\documentclass[11pt,letterpaper]{JHEP3}
\usepackage{array}
\usepackage{cite}
\input epsf.tex
\newcommand{\beq}{\begin{eqnarray}}
\newcommand{\eeq}{\end{eqnarray}}

\def\ltap{\ \raise.3ex\hbox{$<$\kern-.75em\lower1ex\hbox{$\sim$}}\ }
\def\gtap{\ \raise.3ex\hbox{$>$\kern-.75em\lower1ex\hbox{$\sim$}}\ }

\def\be{\begin{equation}}
\def\ee{\end{equation}}
\def\bea{\begin{eqnarray}}
\def\eea{\end{eqnarray}}

\newcommand{\gsim}{ \mathop{}_{\textstyle \sim}^{\textstyle >} }
\newcommand{\lsim}{ \mathop{}_{\textstyle \sim}^{\textstyle <} }
\newcommand{\vev}[1]{ \left\langle {#1} \right\rangle }

\newcommand{\gev}{{\rm GeV}}
\newcommand{\tev}{{\rm TeV}}
\newcommand{\mev}{{\rm MeV}}

\title{A Supersymmetric Twin Higgs}
\author{Spencer Chang$^a$, Lawrence Hall$^{b}$ and Neal Weiner$^a$\\
$^a$ Center for Cosmology and Particle Physics, Dept. of Physics, New York University,
New York, NY 10003 \\
$^b$ Physics Dept. and Theoretical Physics Group, Lawrence Berkeley National Laboratory, University of California, Berkeley, CA 94720\\
\email{sc123@cosmo.nyu.edu, LJHall@lbl.gov, nw32@nyu.edu}
}
\preprint{\today \\ }
\abstract{     
 We present a supersymmetric realization of the twin Higgs mechanism, which cancels off all contributions to the Higgs mass generated above a scale $f$.
Radiative corrections induced by the top quark sector lead to a breaking of the twin sector electroweak symmetry at a scale $f \sim \tev$.  In our sector, below the scale $f$, these radiative corrections from the top quark are present but greatly weakened, naturally allowing a $Z$ boson mass an order of magnitude below $f$, even with a top squark mass of order 1 TeV and a messenger scale near the Planck mass.  A sufficient quartic interaction for our Higgs boson arises from the usual gauge contribution together with a radiative contribution from a heavy top squark.
The mechanism requires the presence of an $SU(2)$-adjoint superfield, and can be simply unified. Naturalness in these theories is usually associated with light winos and sleptons, and is largely independent of the scale of the colored particles.
The assumption of unification naturally predicts the existence of many exotic fields. The theory often has particles which may be stable on collider timescales, including an additional color octet superfield. In the limit that $m_{SUSY} \gg f$, the mechanism yields a UV completion of the non-supersymmetric twin Higgs, and with the notable improvement of a tree-level quartic for the standard model Higgs. In this framework, a successful UV completion requires the existence of new charged fields well below the scale $f$. 
}

\begin{document} 

\section{Introduction}

Supersymmetric extensions of the standard model (SM) tame the quadratic divergences associated with the Higgs boson mass, allowing perfectly natural theories for all energies up to the Planck scale.  Yet at first sight they present a new puzzle:  given all the scalars in the theory, why is it the Higgs boson that acquires a vev rather than a squark or slepton?  Remarkably, radiative corrections to the supersymmetry breaking scalar masses provide a dynamical understanding for why the Higgs, and no other scalar, acquires a mass.  As these mass parameters are scaled to the infrared, they are increased by gauge interactions and decreased by Yukawa interactions.  Thus symmetry breaking is induced for the field that appears in the largest Yukawa coupling but has the weakest gauge interactions.  This field is the Higgs boson, and the resulting Electroweak Symmetry Breaking (EWSB) is driven by the large value of the top quark Yukawa coupling.  

This elegant, almost inevitable, breaking of electroweak symmetry, was seen as a key element of supersymmetric theories in the early 1980s, and, with the precision measurement 
of the weak mixing angle at the beginning of the 1990s, cemented supersymmetry as the leading extension of the Standard Model.  However, the top quark radiative mechanism for electroweak symmetry breaking is considered by many to require fine-tuning --- the problem is that it is simply too efficient, driving too large a Higgs vev so that the $W$ and $Z$ bosons are too heavy.  The size of the negative Higgs boson mass squared, and therefore the size of the EWSB vev, is determined by the top quark Yukawa coupling, the top squark mass and by the number of decades of renormalization evolution.  The top quark is so heavy that the radiative mechanism is extremely powerful: even if the top squark mass is near its experimental limit, scaling from the Planck scale drives too large a Higgs vev.  If gravity mediation of supersymmetry breaking is replaced by gauge mediation at a much lower scale, the experimental limit on the scalar tau mass forces the top squark to be quite heavy, that again the vev is naturally too large.  

The inability of LEP2 to discover a Higgs boson has compounded the problems for top-driven radiative EWSB.  For the Higgs boson to be sufficiently heavy a new quartic Higgs interaction is required beyond that provided by the supersymmetric electroweak gauge interactions.  In the simplest models this can only arise from radiative corrections from a top squark that is significantly heavier than its experimental bound.  This further increases the efficiency of the heavy top radiative EWSB mechanism, leading to significant fine tuning for simple realistic theories. For an excellent discussion of the details, see \cite{Chacko:2005ra}.

This problem of EWSB in supersymmetric theories has received considerably attention recently.  
One can seek an alternative scheme for mediating supersymmetry breaking to the standard model sector at low energies, via a low effective mediation scale and sizable $A_t$ term \cite{Choi:2005hd,Kitano:2005wc}.  The mass correction can also be reduced through conformal dynamics or by generating the top Yukawa at a low scale \cite{Kobayashi:2004pu,Kobayashi:2005mg}.  The Higgs could be a pseudo-Goldstone \cite{Birkedal:2004xi,Berezhiani:2005pb,Roy:2005hg,Csaki:2005fc} or even composite \cite{Harnik:2003rs,Chang:2004db,Delgado:2005fq}, cutting off the log or allowing larger quartics. The Higgs boson could be made heavy by adding additional gauge \cite{Batra:2003nj,Maloney:2004rc} or superpotential \cite{Espinosa:1992hp} interactions.  Furthermore, one might actually have a lighter Higgs, but evade the LEP Higgs bounds by new decay mechanisms \cite{Dermisek:2005ar,Chang:2005ht,Schuster:2005py, Dermisek:2005gg}.

In this paper we introduce a new mechanism that weakens the strength of top-quark radiative EWSB.  It works even for gravity mediated supersymmetry breaking, and does not require the top squark to be light.  It makes use of the twin Higgs mechanism \cite{Chacko:2005pe}.  The standard model has a mirror or twin duplicate, that is guaranteed to have the same couplings by an interchange $Z_2$ parity. The Higgs potential, involving both our Higgs doublet $H$ and the twin Higgs doublet $H'$, possesses an approximate $SU(4)$ symmetry acting on ${\cal H} = (H,H')$. 
A large negative mass squared $-m^2 {\cal H} {\cal H}^\dagger$ leads to a large EWSB in one of the sectors, which by definition is the twin sector.  This was proposed as a way to make progress on the Little Hierarchy Problem in non-supersymmetric theories.  However, in the simplest model a hierarchy of vevs between our sector and the twin sector itself requires some tuning. Nevertheless, adding a twin of the SM does lead to a theory with significantly improved naturalness over the SM, while still preserving agreement with precision electroweak data \cite{Barbieri:2005ri}. The naturalness of the theories has been improved by enlarging the Higgs sector \cite{Chacko:2005vw}, or extending the gauge group to $SU(2)_L \times SU(2)_R$ \cite{Chacko:2005un}. However, in these theories, the cutoff remains quite low. 

Starting with the Minimal Supersymmetric Standard Model (MSSM), we show that adding a twin MSSM$'$ can solve the fine tuning problem of supersymmetry.  There are now two $SU(4)$ Higgs scalars, ${\cal H}_u$ and ${\cal H}_d$, with mass terms $m_u^2 {\cal H}_u {\cal H}_u^\dagger + m_d^2 {\cal H}_d {\cal H}_d^\dagger + B^2({\cal H}_u {\cal H}_d + h.c.)$. If the determinant $m_u^2 m_d^2 - B^4$ is positive there is no EWSB. The top and mirror top radiative corrections reduce $m_u^2$ leading to a negative determinant inducing a large vev for the twin Higgs. The D term quartics explicitly break the approximate $SU(4)$ symmetry, so that a negative Higgs mass squared would also be generated for our sector.  However, employing the ``supersoft" mechanism \cite{Fox:2002bu} on the twin sector, guarantees that this negative Higgs mass squared in our sector is significantly reduced.  Thus the success of top-quark radiative EWSB is restored: its full power is felt only in the twin sector, while in our sector it is still operative, but with a reduced strength.  There is now no barrier to a heavy top squark, needed in some mediation schemes.  In fact, a heavy top squark is now preferred as the simplest origin for an additional quartic interaction to give sufficient mass to the Higgs boson.

\section{A Toy Model}

In non-supersymmetric theories the twin idea may be implemented by having the standard model, SM, an identical mirror, or twin, standard model, SM$'$, and a quartic interaction $H H^\dagger H' H'^\dagger$ such that the combined Higgs potential is approximately $SU(4)$ symmetric.  There is a $Z_2$ symmetry that interchanges the two sectors, forcing $SU(4)$ invariance on the Higgs mass terms, but not on the quartics. A large vev for the twin Higgs breaks $SU(4) \rightarrow SU(3)$ so that our Higgs boson appears as a pseudo-Goldstone boson.  There is an obvious barrier to implementing this in supersymmetry. If our sector is the MSSM and we add an identical twin sector, MSSM$'$, then supersymmetry forbids any quartic interaction coupling our Higgs to the twin Higgs, so that the Higgs quartics cannot be made $SU(4)$ invariant.
A simple way to over come this difficulty is to add a gauge singlet N together with superpotential terms $\lambda N(H_u H_d + H_u' H_d')$.  This leads to a Higgs quartic that couples the two sectors together, and, because it arises from terms in the superpotential that are quadratic in Higgs fields, the $Z_2$ interchange parity is sufficient to guarantee $SU(4)$ invariance.  Hence one sees that the twin idea actually fits very well in supersymmetric theories: the $Z_2$ in the superpotential automatically generates an $SU(4)$ in the quartics, needing no separate assumption.

We begin by considering a toy model, that has the features expected from the twin supersymmetric theory described above, but is simplified in two respects.  Firstly the electroweak gauge group of each sector will be taken to be $U(1)$, and secondly each sector will be taken to have only a single Higgs, rather than separate ones for the up and down sectors. Of course, these do not occur in any realistic supersymmetric model, but they allow a transparent illustration of our mechanism.
Thus we have a $U(1)\otimes U(1)'$ gauge group, with Higgs field ${\cal H}=(h\ h')$ possessing an approximate global $SU(2)$ symmetry, rather than $SU(4)$, and we assume a scalar potential of the form
\be
V= -m^2 {\cal H} {\cal H}^\dagger + \frac{\lambda^2}{2} ( {\cal H} {\cal H}^\dagger )^2 +\delta m^2 h h^\dagger + \frac{g^2}{2} (h h^\dagger)^2 + \frac{g'^2}{2} (h' h'^\dagger )^2.
\ee
Each term in the potential has an important significance. The negative $SU(2)$ invariant mass term arises from the large radiative correction in the top sector, and will be the origin of EWSB, both in the twin sector and in our sector. We imagine that if this negative mass squared were in our sector alone it would lead to the $Z$ boson being to heavy, and our aim is to understand how the presence of the twin sector could reduce the natural value for the $Z$ boson mass.  The quartic coupling $\lambda^2$ 
is the toy model version of the $SU(4)$ invariant Higgs quartic that comes from the interaction with the singlet superfield $N$.  The quartic interactions proportional to $g^2$ and $g'^2$ are the toy analogue of the electroweak $D^2$ terms in our sector and the twin sector.  The $Z_2$ parity sets $g' = g$, but we keep the prime in the twin case so that we can keep track of the origin of the two interactions.  These gauge quartics explicitly break the global $SU(2)$ symmetry of the toy model, and we shall return to this point shortly. Finally, we have allowed for a small $SU(2)$ and $Z_2$ symmetry breaking mass term, $\delta m^2$.

Following the twin idea, suppose that the large negative mass squared leads to EWSB in the twin sector, then
\be
\vev{h'^2} = \frac{m^2}{g'^2 + \lambda^2}.
\label{h'vev}
\ee
Upon integrating out the heavy $h'$, we are left with a low energy effective theory for $h$, with potential
\be
V=\left(\delta m^2-\frac{g'^2}{g'^2 + \lambda^2} m^2\right) h h^\dagger + \frac{1}{2} \left( g^2 +\frac{\lambda^2}{g'^2+\lambda^2}g'^2 \right)( h h^\dagger)^2.
\ee
In the limit that the $SU(2)$ violating terms $\delta m^2$, $g^2$ and $g'^2$ go to zero, we have an exact $SU(2)$ symmetry and $h$ is a Goldstone boson. In the presence of the $SU(2)$-violating terms, the quartic interactions for $h$ are welcome since we need them to get the Higgs boson sufficiently heavy.  However, the mass terms are problemmatic if they are two large. The contribution from $\delta m^2$ can be naturally small, because the $Z_2: h\leftrightarrow h'$ exchange symmetry enforces an accidental $SU(2)$ symmetry on the quadratic terms. However, the $SU(4)$-violating D-term quartics are inevitable in supersymmetry, so reducing or eliminating the contribution to $h h^\dagger$ proportional to $g'^2$ is the most significant challenge.  This term arises because the $g'$ quartic leads to a shift in the vev away from the $SU(2)$ invariant value, as shown in eq. \ref{h'vev}. How can this effect be eliminated?

\subsection*{Modifying the toy model}
The troublesome quartic interaction arising from the twin sector electroweak $D$ terms can be removed by the inclusion of a new superfield.  The analysis below in the toy model corresponds to the inclusion of a ``supersoft'' supersymmetry breaking term \cite{Fox:2002bu} in the realistic theory, as discussed in the next section.

We extend the toy model by including a real singlet field $s$, with a potential
\be
V_s = \frac{\delta m_s^2}{2} s^2 + \frac{1}{2}\left( m_s s + g' h' h'^\dagger\right)^2.
\ee
Such a potential can arise naturally in the presence of D-term supersymmetry breaking \cite{Fox:2002bu}. When $h'$ acquires a vev, a tadpole for $s$ is induced so that it is convenient to redefine $s$ by
\be
s \rightarrow s- \frac{g' m_s}{m_s^2 + \delta m_s^2} h' h'^\dagger.
\ee
This new field is not canonically normalized, so if we want to study the masses of $s$ or $h'$, we should shift back to canonically normalized fields. However, for the purposes of studying vevs and determining the properties of the pseudo-Goldstone boson, this is adequate.

Notice that with such a potential, in the limit that $\delta m_s^2 \rightarrow 0$, such a redefinition removes the $SU(4)$ violating quartic $g'^2$. At leading order in $\delta m_s^2$ and $\delta m^2$, the potential for $h$ is now
\be
V= \left(\frac{\delta m^2}{m^2}-\frac{g'^2 \delta m_s^2}{\lambda m_s^2} \right) m^2 \, h h^\dagger + \left(\frac{g^2}{2} + \frac{g'^2 \delta m_s^2}{2 m_s^2}\right) (h h^\dagger)^2.
\ee
Therefore we have reduced the problem of removing the problematic $g'$ term to a problem of keeping certain supersymmetry breaking masses small. 

In the realistic supersymmetric model, we will need two fields, a singlet $S$ and an $SU(2)$ triplet $T$, which will pick up radiative corrections to their masses. Consequently, the naturalness of the models will be related to the size of SUSY breaking masses for {\em electroweak}-charged superpartners (specifically winos), rather than for colored superpartners, such as gluinos.

\section{A Supersymmetric Twin Higgs}
Let us begin by taking the MSSM and creating a twin copy, the MSSM$'$. We will refer to these sectors as the visible and twin sectors, respectively. We will insist upon a $Z_2:\ MSSM\leftrightarrow MSSM'$ symmetry. The general spirit of the construction will be this: all supersymmetric couplings will respect the $Z_2$, and all soft supersymmetry breaking operators (terms that are log sensitive)  will also respect the $Z_2$. In principle, we can include supersymmetric $\mu$-type masses which violate the $Z_2$, but they are unnecessary phenomenologically. The origin of the $Z_2$ breaking will be a hidden sector $D$-term, which will generate supersoft supersymmetry breaking terms only in the twin sector. We will return to this point in a moment.\footnote{An alternative approach, employing an $SU(2)_L\times SU(2)_R$ gauge group with no twin sector, has been explored in \cite{martin}.}

In order to establish $SU(4)$ invariant quartics, we will employ the superpotential
\be
W=\lambda N {\cal H}_u {\cal H}_d
\label{eq:quartic}
\ee
where ${\cal H}_i=(H_i\ H'_i)$ and $N$ is a singlet superfield under both the MSSM and MSSM' gauge groups. 
Notice that because the superpotential is bilinear in the Higgs fields, the $Z_2$ symmetry alone is sufficient to achieve an $SU(4)$ in the quartic induced in the potential from the $F_N^2$ term. 

Of course, we still have the $SU(4)$ violating D-term quartics as in the toy model, which we must cancel in the twin sector. This is where we will employ the $D$-term supersymmetry breaking operators. We add a singlet $S'$ and a triplet $T'$ under the MSSM$'$ allowing us to expand our superpotential to
\be
W=\lambda N {\cal H}_u {\cal H}_d + \frac{W^D_\alpha}{\Lambda_{S'}} W^{'\alpha}_Y S' +\frac{W^D_\alpha}{\Lambda_{T'}} W^{'\alpha}_{SU(2)} T'.
\ee
To simplify our calculations, we will add a large soft mass for $N$, such that $\vev{N}=0$ and we can decouple it. 
$W^D_\alpha$ is a spurion, reflecting the $D$-term of some hidden sector $U(1)$, so $\vev{W^D_\alpha}= \theta_\alpha D$. The effect of these operators has been explored elsewhere \cite{Fox:2002bu}. They generate scalar masses $m_S$ and $m_T$, with related trilinears which we describe below. They also generate a Dirac mass between the fermion and the gaugino of size $m_S/2$ and $m_T/2$.

At this point we have not included comparable operators for the MSSM, which breaks the $Z_2$. However, because these are {\em supersoft} SUSY breaking operators, that is, they induce no corrections to the RG flow of the soft SUSY breaking masses, all contributions to the $SU(4)$ violations will be loop suppressed, with no log enhancement. (A triplet under the MSSM $SU(2)$ must be added in order to preserve the $Z_2$ of the gauge couplings, but no equivalent superpotential term need be added.)

Aside from these small supersoft effects, the remainder of the Higgs potential arises from $SU(4)$ (i.e., $Z_2$) preserving soft masses $m_{h_u}^2$, $m_{h_d}^2$and $B^2 {\cal H}_u {\cal H}_d + h.c.$. It is important to note that $m_{h_u}^2,m_{h_d}^2$ must both be positive as the quartic in eq. \ref{eq:quartic} will not stabilize against breaking in these directions. The proper spectrum is $m_{h_d}^2 > B^2 > m_{h_u}^2$. However, this is a natural expectation in that the top/stop loops will drive down $m_{h_u}^2$, turning the determinant of the mass matrix negative at a low scale.
  
Let us go to the basis ${\cal H}=\sin\beta {\cal H}_u + \cos\beta {\cal H}_d$, $\bar{\cal H}=\sin\beta {\cal H}_d-\cos\beta {\cal H}_u$ where the mass matrix is diagonal and of the form
\be
\pmatrix{-m^2 & 0 \cr 0 & M^2}
\ee
Here we know that only the field $\cal H$ will acquire a vev, so we can set $\bar {\cal H}=0$. Then potential reads
\bea  \nonumber
V=&-&m^2 {\cal H} {\cal H}^\dagger +\frac{\lambda^2}{4} ({\cal H}{\cal H}^\dagger)^2 \sin^2 2 \beta+ \frac{g^2+g_Y^2}{8}(H H^\dagger \cos (2 \beta))^2 \\ &+& \frac{1}{2}\left( m_S S' + g_Y' H' H'^\dagger \cos(2 \beta) \right)^2 + \frac{1}{2}\left( m_T T' + g' H' H'^\dagger \cos(2 \beta) \right)^2 \\ \nonumber &+&\frac{\delta_S^2}{2}S'^2 + \frac{\delta_T^2}{2} T'^2
\eea
where the $S'$ and $T'$ fields are in an abuse of notation the real parts of the respective scalars.
As before, we can redefine
\be
S' \rightarrow S' - \frac{g'_Y m_S H'^2 \cos(2 \beta)}{m_S^2 + \delta_S^2} \hskip .5 in T' \rightarrow T' - \frac{g' m_T H'^2 \cos(2 \beta)}{m_T^2 + \delta_T^2} 
\ee
These redefined fields will not acquire vevs, so we can set them to zero in the potential, leaving us with
\bea  \nonumber
V=&-&m^2 {\cal H} {\cal H}^\dagger +\frac{\lambda^2}{4} ({\cal H}{\cal H}^\dagger)^2 \sin^2 2 \beta+ \frac{g^2+g_Y^2}{8}(H H^\dagger \cos (2 \beta))^2 \\&+& \frac{\gamma g^{'2}+\gamma_Y g^{'2}_Y}{8}(H' H'^\dagger \cos (2 \beta))^2 
\eea
where $\gamma_Y = \delta_S^2/(m_S^2 + \delta_S^2)$ and $\gamma = \delta_T^2/(m_T^2 + \delta_T^2)$. To the extent that the corrections $\delta_{S,T}$ are small compared to $m_{S,T}$ these will be small numbers.

Now $H'$ will acquire a vev 
\be
\vev{H^{'2}}=f^2 = \frac{4 m^2}{g^{'2} \gamma \cos^2 (2 \beta) + g_Y^{'2} \gamma_Y \cos^2 (2 \beta) + 2 \lambda^2 \sin^2 (2 \beta)}
\ee
That the vev is in the twin direction is a dynamical selection due to the smaller quartic, not a choice.
Integrating out the $H'$ and taking $g',g_Y'=g,g_Y$ gives us
\be
V= -m^2\frac{(\gamma g^2 + \gamma_Y g_Y^2) \cot^2 (2 \beta) H^2}{2 \lambda^2} +\frac{(1+\gamma)g^2 + (1+\gamma_Y) g_Y^2}{8} H^4 \cos^2 (2 \beta)
\ee
This is the principal result of this paper. We have an order one quartic, but with a tree level mass suppressed by small numbers $\gamma,\ \gamma_Y$.

\subsection{Scales and limits}
Now we have established the basic tools for constructing a twin Higgs. But is it a twin Higgs theory, or is it a supersymmetric theory? The answer depends on the scales $m_{SUSY}$ and $f$.

In the limit that $m_{SUSY}/f \ll 1$, the theory is nearly a standard supersymmetric theory. Because we have invoked supersymmetry breaking masses to break the $SU(4)$, one has a supersymmetric model in the decoupling limit, with the Higgs fields $H,A$ with masses $O(f)$. As a consequence, the quadratic divergence associated with the Higgs self-coupling is not cancelled until the scale $f$. However, the more serious quadratic divergences, associated with the top Yukawa and gauge couplings, are cancelled by supersymmetry, and the logarithmic divergence cut off at $f$. All other SUSY masses will arise from higher scales and will, in general, be much larger than $m_h$. 

Since SUSY is invoked in the breaking of $SU(4)$, and because the heavy Higgses appear at the scale $f$, it is clear that the scale of $SU(4)$ breaking cannot be much higher than the SUSY scale.\footnote{This could be possible if there is a linear term for $N$ of size $f^2$ in the superpotential.  We do not consider this possibility due to the unknown mechanism which chooses this scale, but it is an interesting model in it's own right as a solution to the Little Hierarchy.}

In the limit that $m_{SUSY}/f \gg 1$ we achieve a standard twin Higgs model, albeit one greatly improved over the model presented in \cite{Chacko:2005pe}, because of the presence of a tree-level quartic coupling. In this case, it is the $SU(4)$ which is protecting the Higgs mass, and supersymmetry + $Z_2$ protecting the $SU(4)$. 

Since SUSY is invoked to protect the $SU(4)$, it is clear that one cannot take the SUSY breaking scale arbitrarily high. In particular, we shall see that the winos, and the sleptons in general,  must remain light in order for the cancellation of the twin quartic to occur.

Ultimately, we are directed towards the $m_{SUSY}\sim f$ region. In this region, there are no quadratic divergences above $f$, but the Higgs mass is cancelled up to one loop corrections by the $SU(4)$ breaking. In general, some fields (namely, squarks and gluinos) will typically appear above the scale $f$, while others (sleptons, winos) will typically appear below the scale $f$. The particular limits, however, require a more careful study of the radiative corrections to the theory.

\subsection{Radiative corrections}
We now understand that it is possible to have a small soft mass for the Higgs field at the scale $f$, but we have translated the problem of the Higgs mass divergences into a problem of controlling the masses of $S$ and $T$. This is much easier because we lack the sizeable Yukawas to colored particles that are so problematic in the MSSM. Nonetheless, radiative corrections impose significant constraints on the spectrum.

Because $S$ has no gauge charges and no Yukawa couplings, it has no radiative corrections to its mass. However, $T$ receives corrections to its mass from a variety of sources. Loosely, we can group the radiative corrections into two parts: the log-enhanced RG flow, and the finite one-loop effects arising from the $Z_2$ breaking supersoft terms.

The RG flow is simply the usual contribution (see, e.g., \cite{Martin:1997ns})
\be
\frac{\partial m_T^2}{\partial \log \mu} = -\frac{2 C_2^T \alpha_2}{\pi} M_2^2.
\ee
where $M_2$ refers to the $Z_2$ preserving Majorana mass for the $SU(2)$-gauginos. There are also two-loop corrections to its mass from the soft masses of other fields charged under $SU(2)$ (see, e.g., \cite{Arkani-Hamed:1998kj})
\be
\frac{\partial m_T^2}{\partial \log \mu} = \frac{ C_2^T \alpha_2^2}{2\pi^2}\sum_i T_i m_i^2,
\ee
where $T_i$ is the Dynkin index of the $i$ representation.
Both of these terms should be properly evolved in a given model, as they depend on the value of $\alpha_2(\mu)$, which, in turn, depends on the particle content of the theory. However, given that new matter makes UV values of $\alpha$ larger, we can use the MSSM limit for the former as an estimate of the radiative correction assuming the masses are generated at a high scale,
\be
\delta m_T^2 \sim 0.5 C_2^T M_2^2
\ee
In contrast, the effects of the two-loop running cannot be simply estimated without input as to the values of the scalar masses at the high scale as well. However, they are in general much smaller than the one loop contribution from gaugino masses. 

Additionally, we must concern ourselves with the $Z_2$-violating corrections arising from the Dirac gaugino masses in the twin sector. In the limit that the Dirac masses are much larger than the Majorana masses, the radiative corrections are simply given by \cite{Fox:2002bu}
\be
\delta m^2_{\tilde{x}} = \frac{C_i(\tilde{x}) \alpha_i m_{i}^2}{4\pi} \log(4)
\label{eq:ssoftcorr}
\ee
where $m_i$ is the supersoft mass of the scalar associated with the gauge group indexed by $i$ (as a warning, throughout the rest of the text, we will refer to these as $m_S, m_T, m_O$).  The $\log(4)$ term arises from the ratio of the scalar and Dirac gaugino masses squared. When the Dirac and Majorana masses are comparable, there is unfortunately no simple formula, but the overall magnitude of the effect does not change.

\subsection{Naturalness \label{sec:naturalness}}
As in the MSSM, many questions of naturalness arise due to the assumption of unification. If experimental limits on charginos or staus, for instance, indirectly imply scales for gluinos and squarks, there can be a significant effect on the naturalness of the theory. However, in the case of the supersymmetric twin Higgs, there are significant direct relationships between the wino Majorana mass, and the Dirac bino and wino masses.

In particular, we require that the soft mass for $T$ be small in order that the cancellations of the D-term quartic in the twin sector are complete. If this is not the case, one ends up with a correction to the Higgs mass $O(\delta m_T^2 m_{\cal H}^2/m_T^2)$. If one wishes to cancel the existing mass term of the Higgs to a few percent, and given that $\delta m_T \sim M_2$, one must have $m_T \approx 5-10 \times M_2$. Given limits on charginos from LEP2, this tells us $m_T \gsim 500\ \gev \sim 1\ \tev$.

It is important to state that most, but not all, models will generate scalar masses at a similar scale to gaugino masses. Nonetheless, some models, e.g., low scale gaugino mediation \cite{Kaplan:1999ac,Chacko:1999mi}, could have the scalar masses considerably lower than the gaugino masses. Thus, while we believe that a lighter wino is a generic feature of these theories, it is by no means essential.

A similar statement can be made about the scalar, specifically slepton, masses. While they can, in principle, be quite heavy, with only two-loop running feeding into the $T$ mass, they are quite often generated from gauge interactions, and so we expect those masses to be comparable to the $T$ mass as well. A good estimate of the ratio of masses squared would be the ratio of the casimirs of the fundamental to the adjoint representation. Hence, a $T$ lighter than $400\ \gev$ would likely be accompanied by left-handed sleptons in the $250\ \gev$ range.  Again, these limits are not absolute, but demonstrate that preserving the accidental $SU(4)$ is associated with light ($m<f$) fields, although such a statement is not obvious from the outset.

The Dirac masses in the twin sector are more model-independent in their effects. In fact, there are upper bounds on their size due to naturalness.  The presence of a large supersoft $m_T$ will generate radiative corrections in the twin sector that are not cancelled in the visible sector. From eq. \ref{eq:ssoftcorr}, we see that there will be a correction roughly $\delta m_h^2 \sim {\rm few}\ \times 10^{-3} m_T^2$. For a natural completely natural Higgs mass ($O(100 \gev)$), one expects $m_T \lsim 2 \tev$, while for moderate tuning, $m_T \lsim 5 \tev$ would be reasonable.

The combination of these two requirements suggests that a light wino in the visible sector is most natural. While there is no absolute upper bound, the requirement of naturalness suggests an upper bound of $m_{\rm wino} \lsim 400 \gev$ in the most natural models. Sleptons are expected to be light, but, similarly, the limits are not absolute.  Once unification is included, our expected parameter range will narrow somewhat.  

To summarize, there is in general a correlation between the soft contributions to the $T'$ mass and the electroweak soft masses for the wino and sleptons.  When considering experimental limits on the visible particles, requiring that this doesn't upset the D-term cancellation gives a lower bound for the supersoft $T'$ mass.  On the other hand, there is an upper bound on the same mass due to naturalness, as the $Z_2$ violation becomes too large.  Therefore, these considerations will impact prospects of collider searches.    

\section{Unification \label{sec:unification}}
The inclusion of an $SU(2)'$ adjoint in the twin sector necessitates one in the visible sector by $Z_2$. The natural consequence of this is to spoil unification. However, this is easily addressed by GUT-completing the adjoint with additional fields.

There are essentially two options, as outlined in \cite{Fox:2002bu}. The most obvious would be to GUT-complete into a {\bf 24} of $SU(5)$. This amounts to adding a total of five flavors to each sector, resulting in a landau pole below the GUT scale. It is difficult to continue to claim the quantitative successes of unification under such circumstances.

A more restrained approach is to GUT-complete into a {\bf 24} of $SU(3)^3$. This amounts to the addition of an {\bf 8} of $SU(3)$ color, as well as a vectorlike pair of $(1,2,\pm 1/2)$ fields, two pairs of $(1,1,\pm 1)$ fields, and four gauge singlets. This amounts to the addition of 3 flavors to the theory, which retains perturbativity in the theory, and thus the quantitative successes of unification.

The basic features of the spectrum have been laid out in \cite{Fox:2002bu} which we follow here. The $\beta$ functions are given by $(b_1,b_2,b_3)=(33/5,1,-3)$ in the MSSM. With the additional matter, we now have $(48/5,4,0)$. Since $\alpha_3$ is asymptotically flat at one loop, we can take $\alpha_i (M_{GUT}) = \alpha_3$. The supersoft mass parameters run due to gauge kinetic term renormalization, as well as the adjoint kinetic term renormalization. Hence
\bea
&m_S& = \left(\frac{\alpha_1(\mu)}{\alpha_1(M_{GUT})}\right)^{1/2} M
\nonumber
\\
&m_T& = M
\\
\nonumber
&m_O& =  \left(\frac{M_{GUT}}{\mu}\right)^{\frac{3 \alpha_3}{2 \pi} }M=M^\frac{2\pi-3 \alpha_3}{2\pi} M_{GUT}^\frac{3\alpha_3}{2\pi}
\eea
where $m_O$ is the supersoft mass of the new scalar color octet, and $M$ is the common supersoft scalar mass at the unification scale.\footnote{It is important to note that these expressions are only true under the assumption of no additional matter in the UV. In a gauge mediated scenario, for example, the additional messengers would modify the expressions above.}

In terms of the unified mass term, the one loop scalar soft masses squared are (again, see \cite{Fox:2002bu})
\bea
&m_r^2& = \frac{C_1(\phi)\alpha_1^2 (\mu) M^2 \log(4)}{4 \alpha_1(M_{GUT}) \pi}
\nonumber
\\
&m_l^2& = \frac{C_2(\phi)\alpha_2 (\mu) M^2 \log(4)}{ 4\pi}
\\
&m_c^2&= \frac{C_3(\phi)\alpha_3 (\mu) M^2 \log(4)}{4 \pi}\left(\frac{M_{GUT}}{m_3}\right)^{\frac{3\alpha_3}{\pi}}
\nonumber
\eea
Because the color octet mass is so large, these corrections to the twin squarks will induce two-loop corrections to the twin Higgs mass, which are $Z_2$ violating. At leading order in $\alpha_3$, this is given by
\be
\delta m_h^{'2} 
= - \frac{ \lambda_t^2 \alpha_3 \log(4)}{2 \pi^3} \left(\frac{M_{GUT}}{m_3}\right)^\frac{3 \alpha_3}{4\pi} \log(M_3/m_{\tilde t'}) 
\ee

Ultimately, the tension is between our desire to have a large $m_T$, and hence a robust cancellation of the $D$-term quartics and the consequent radiative correction to the twin Higgs mass squared. Because in this unified scenario, the two loop $Z_2$-violating contributions are larger than the one loop $Z_2$-violations, the upper bound on $m_T$ comes indirectly from an upper bound on $m_O$. Numerically, when one includes the relationship to $m_O$, one finds $\delta m_h^2/m_T^2 \approx 10^{-2}$, so a completely natural model with $m_h^2 \sim (100 \gev)^2$ would require $m_T \sim 1\ \tev$. The upshot of this are slightly more stringent requirements on the wino and slepton masses in their relation to the radiative corrections to the soft mass of $T$.

Having assessed the consequences for naturalness on MSSM fields, where are the new exotic fields in the observable sector? They must have some mass, which is most simply understood by adding a supersymmetric
 $\mu$-term masses for these fields. A priori, this mass needn't be the same as the mass in the twin sector, in that it will not lead to any sizeable $Z_2$ violating radiative corrections. Thus, in principle, the new fields in the visible sector can be quite heavy ($\sim \tev$).
 
However, under the assumption that only supersoft violates the $Z_2$, we can make stronger statements. Because the supersymmetric masses interfere with the cancellation of the D-term in the twin sector, the masses for $S$ and $T$ should be small - at most of the order of the wino mass. Under the assumption of unification, we can calculate the spectrum in the visible sector.
 \bea
 \mu_S &=& \mu \nonumber \\ 
 \mu_T &=& \frac{\alpha_{GUT}}{\alpha_2} \mu \\
 \mu_O &=& M_{GUT}^{\frac{6 \alpha_3}{2 \pi}} \mu^{\frac{2 \pi - 6 \alpha_3}{2 \pi}} \nonumber 
 \eea
 
Numerically, we have $\mu_S:\mu_T:\mu_O \approx 1:3:21$. If we assume $Z_2$ symmetry in these masses, and require $\mu_T \lsim 200 \,\gev$, then we have $m_O \lsim 1.5 \,\tev$. Such a particle should be produced at the LHC. We will discuss the phenomenology in section \ref{sec:pheno}.

 The additional $SU(2)\times U(1)$ fields which complete the $SU(3)^3$ adjoint (the so-called ``bachelor'' fields \cite{Fox:2002bu}), should have a mass related by unification to these $\mu$-terms. However, the RG-evolution of these masses depends on the particular couplings of the bachelor fields, specifically Yukawa-type couplings. As we shall see in section \ref{sec:cosmo}, such couplings may naturally enable a Froggatt-Nielsen theory of flavor \cite{Froggatt:1978nt}. Hence, we can say little about their specific spectrum except that they should be in the $100 \ \gev$ range rather than the $1 \ \tev$ range.

\subsection*{$\Lambda_{QCD}'$}
Under the assumption of unification, the new colored fields in both sectors can affect $\Lambda_{QCD}$. Under the assumption that the same sets of fields remain light, and so contribute to the running of the strong coupling (i.e., gluons, up, down and strange quarks), there is a simple relationship between $\Lambda_{QCD}'$ and $\Lambda_{QCD}$, specifically
\be
\Lambda_{QCD}' = \Lambda_{QCD}\ \Pi_i \left(\frac{M_i'}{M_i}\right)^{-b_i/b_{light}}
\ee
where $M_i$ ($M_i'$) is the common mass of fields in the observable (twin) sector, contributing a {\em positive} value $b_i$ to the $\beta$ function of QCD, while $b_{light}$ is the {\em negative} contribution to the $\beta$ function from the fields lighter than the strong coupling scale.

Assuming common squark masses, the twin strong couplings scale is simply
\be
\Lambda_{QCD}' = \Lambda_{QCD} 
\left({f \over v} \right)^{2/9}
\left({m_{\tilde q}' \over m_{\tilde q}} \right)^{2/9}
\left({m_{gluino}' \over m_{gluino}} \right)^{2/9}
\left({m_{ferm}' \over m_{ferm}} \right)^{2/9}
\left({m_{scal}' \over m_{scal}} \right)^{1/18}
\left({m_{pseudo}' \over m_{pseudo}} \right)^{1/18}
\ee
where $m_{ferm}$, $m_{scal}$ and $m_{pseudo}$ are the masses of the octet fields. If we want TeV squarks in the observable sector, we cannot have twin squarks much heavier than 3 TeV without generating unacceptably large two-loop corrections to the Higgs mass. The gluino and scalar octet masses could be a factor of ten larger, while the pseudoscalar, picking up its $Z_2$ violating mass difference from the radiative corrections, is likely of a similar ratio to the squarks. 
Thus, we expect the twin QCD scale to be roughly $\Lambda_{QCD}' \simeq (4 - 7)\times \Lambda_{QCD}$.
\section{Cosmology \label{sec:cosmo}}

As discussed previously \cite{Chacko:2005pe, Barbieri:2005ri}, the presence of a twin sector can have significant cosmological effects if the extra degrees of freedom were in thermal equilibrium at some earlier time in the universe.  In Twin Higgs models, the presence of cross term quartic couplings that link twin sector Higgses to visible ones mediate processes that keep the twin sector in thermal equilibrium.  Below the scale of the two electroweak breakings, one can integrate out the massive scalars to generate four fermion couplings between the two sectors of the following form:
\be
\frac{m_\Psi \, m_{\Psi'}}{m_h^2 \, f^2} \bar{\Psi}\Psi \, \bar{\Psi}' \Psi'
\ee  
These operators keep the sectors in thermal equilibrium to low temperatures; for example for an f scale of 500 GeV, processes that convert charm quarks into twin muons occur at a rate
\be
\tau^{-1} \sim \frac{m_{\mu'}^2 m_{c}^2}{m_h^4\, f^4}\, T^5 
\ee
using $m_c = 1.4 \,\gev, m_\mu' = 300 \,\mev$ determines that this process decouples at a temperature of about 2 GeV.  Other processes give similar decoupling temperatures.

If one insists on a low reheat temperature, anywhere from a few MeV to just above the QCD phase transition, cosmological difficulties are evaded.  However, in more standard thermal histories, one is forced to address the issues raised by the additional thermal degrees of freedom.  Of primary concern is additional relativistic energy at the BBN era as well as its effects on the CMBR.  The first assumption would be that the $\gamma'$ is massless and the twin neutrinos are a factor of $f/v$ or $(f/v)^2$ heavier than in the Standard Model - assuming no other $Z_2$ violation is present in the theory.  If this is the case, the crucial factor is the temperature of these light degrees of freedom, when the regular universe is at MeV temperatures.  The additional relativistic energy density is given by 
\be
\rho' = \rho_{1\nu}\left[3\left(\frac{T_{\nu'}}{T_\nu}\right)^4 + \frac{8}{7}\left(\frac{T_{\gamma'}}{T_\nu}\right)^4\right] 
\ee
when normalized to the relativistic energy of one SM neutrino.  Since current BBN limits restrict the number of additional neutrinos to be no more than one (for a recent discussion, see \cite{Dolgov:2003sg}), the bracketed terms are constrained to sum to less than one.  There are two relevant scenarios for which we can determine the temperatures involved.  First, if the twin electrons are below the decoupling temperature, when they annihilate, the $T_{\gamma'}$ will be enhanced relative to $T_{\nu'}$ as in the SM.  In this case, we have 
\be
\frac{1}{1.4}\; \frac{T_{\gamma'}}{T_\nu} = \frac{T_\nu'}{T_\nu} = \left(\frac{g'_*(T_d)}{g_*(T_d)} \right)^{\frac{1}{3}}.
\ee
where $T_d$ is the decoupling temperature and $g_*(T_d)$ and $g'_*(T_d)$ the number of relativistic degrees of freedom (including 7/8 for fermions) in the visible and twin sectors respectively at decoupling.  The BBN constraints requires that $g'_*(T_d)/g_*(T_d) \lesssim 1/4.5$, which is a reasonably stringent constraint.  The other relevant case is when only the $\gamma'$ and $\nu'$ are below the decoupling temperature.  Then there is no relative heating up of the photons and we have 
\be
\frac{T_{\gamma'}}{T_\nu} = \frac{T_\nu'}{T_\nu} = \left(\frac{10.75}{g_*(T_d)} \right)^{\frac{1}{3}}.
\ee
In this case, the BBN constraint requires $g_*(T_d) \gtrsim 32,$ or equivalently $T_d \gtrsim T_{QCD}$, just above the QCD phase transition and consistent with the numbers given above.

To be consistent with the constraints in either case requires a smaller $g'_*(T_d)$ than that given by a spectrum scaled by $f/v$.  In the first case, one requires at least the second generation to be heavier than the 1-2 GeV decoupling temperature (giving $\Delta N_\nu \sim 1$) whereas in the second case, the first two generations must be above the decoupling temperature.  This required increase in the prime Yukawas can come from different realizations.      


The first possibility is to just allow $Z_2$ breaking in the Yukawas for the first two generations as in \cite{Chacko:2005pe,Barbieri:2005ri}.  They can be increased to a sufficient level without introducing significant $Z_2$ breaking in the Higgs potential.  This is potentially given by some high scale flavor physics, but appears to be an ad hoc assumption that runs counter to the $Z_2$ symmetry of the model.  On the other hand, the supersoft $Z_2$ breaking already yields what appears to be hard $SU(4)$ breaking in the Higgs potentials at low energies. Given this, it is not surprising that other indirect effects of the supersoft $Z_2$ breaking can induce what appear to be hard breaking in other areas of the theory. 

This can be trivially implemented using the large vevs of the adjoint fields in the context of a low-scale Froggatt-Nielsen model \cite{Froggatt:1978nt}. Most simply, if the Yukawas of the twin and visible sectors are generated by the presence of interactions of new fields in the $100\, \tev$ range, then it is quite plausible to imagine that these Yukawas would depend on the vevs of the singlet fields as
\be
\frac{Q'U'S'H_u'}{M} + \frac{Q'D'S'H_d'}{M}+\frac{L'E'S'H_d'}{M}
\ee
or
\be
\frac{Q U H_u}{M+S'} + \frac{Q D H_d}{M+S'}+\frac{LEH_d}{M+S'}
\ee
In the former case, the singlet vev would generate larger Yukawas in the twin sector, while in the latter case the singlet vev would suppress Yukawas in the visible sector.\footnote{Froggatt-Nielsen models are easily thought of as a mixing to a vectorlike fermion which is integrated out. For instance a superpotential $m \psi^c H + M \psi \psi^c + y \chi \chi^c \psi$ generates a low energy interaction $-y \frac{m}{M} \chi \chi^c H$. In our theory, if we simply allow either $m$ or $M$ to depend on $\vev{S'}$, we trivially achieve either of the examples above.  Note that the relevant new fields exist in trinification, by using the electroweak doublets in the ESPs as the $\psi, \psi^c$.  Therefore, the unification story given in section \ref{sec:unification} can still be maintained.  In addition, for this choice, there is a GIM mechanism at work for the first two generations; thus the scale $M$ can be lowered to almost the weak scale.}  It is also important to emphasize that the constraints do not require a significant change in the Yukawas, since raising the 2nd generation above a few GeV is at most an order of magnitude change in the Yukawas.  

The particular model which achieves this is not important for our present discussion. However, the necessity of this (relatively) low scale of flavor suggests that, in spite of the large SUSY breaking masses in the theory, flavor violation may still be large enough to be observable.  Similarly, even though the BBN constraints can be satisfied, improved measurements on relativistic energy density at BBN or in the CMB are expected to give deviations from that of the SM.

\section{Phenomenology}
\label{sec:pheno}
The general twin Higgs phenomenology applies to this model, for instance there are new invisible Higgs decays into twin particles \cite{Chacko:2005pe,Schabinger:2005ei}.  This is mediated due to mixing and hence gives an $O((v/f)^2)$ branching ratio into invisible decays. However, since the model is also supersymmetric, there are new signatures of the model within the supersymmetry phenomenology.  Furthermore, since this is the primary signal of new physics, one can hope to find features (or even smoking guns) that indirectly confirm this model at future colliders. 

For Higgs physics, we expect a MSSM two Higgs doublet model in the decoupling limit.  This is due to the fact that there is only a single Higgs doublet that is protected to be light by the Twin Higgs mechanism.  Also, the SM Higgs mass limit suggests that $\tan \beta$ is reasonably large.  So although this is not a smoking gun, if such Higgs physics is not observed, this model will be disfavored.  Similar statements can be made within the top squark sector.  They should be reasonably heavy so as to make the Higgs heavy enough, probably in a region that would be considered tuned normally.  On the other hand, if there is gaugino unification, the gluino is probably lighter than expected, since the gaugino spectrum is light (see gaugino comments below).    

However, there are interesting constraints on the electroweak sector within the supersymmetry breaking sector. As discussed in section \ref{sec:naturalness}, the cancellation of the D-term in the twin sector requires the additional masses for the electroweak adjoint scalars to be small, which implies that the common supersymmetry breaking masses for electroweak particles are also small.  Within the gaugino/higgsino sector, this implies that they are light and should be below roughly $400 \, \gev$. If these are light enough, there will be excellent prospects for their direct detection at LHC.  This also applies to the slepton sector, not only giving hope for their direct production, but also suggesting that decay cascades of colored sparticles should end up with copious leptons. 

While the scale of leptons and winos is closely related to the scale of electroweak symmetry breaking, ironically, the Higgsino mass parameter $\mu$ is essentially decoupled from it. Because the contribution from $\mu$ is cancelled off by the twin sector (under the assumption of $Z_2$ symmetry), it does not contribute significantly to the final scale for $m_Z$. In usual SUSY scenarios, $\mu$ often ends up quite large, decoupling the Higgsinos from the other electroweak fields. Here, this is not a requirement. Additionally, because we require all other $\mu$-type masses to be small, it is likely that Higgsinos will be light as well.

The model also predicts the existence of exotics in the form of the adjoints and, including unification, their GUT partners, the bachelors.  One would also expect the electroweak adjoints to be light, however there can be a supersymmetry preserving mass (a soft $Z_2$ breaking) for the visible adjoints and bachelors that makes them heavier.  So although their discovery prospects are not linked to naturalness, finding these exotics would be an intriguing hint of this model.

There is also potential new physics for Dark Matter and, in general, long lived particles in the model. Because of possibly light Higgsinos, annihilation in the early universe can be more efficient, bringing down the relic abundance without resorting to coannihilation or s-channel poles. There is both the possibility of twin nucleons comprising a large portion of the Dark Matter as well as the conventional LSP.\footnote{Due to the $\lambda N {\cal{H}}_u {\cal{H}}_d$ coupling, there is only a single R-parity and hence only one LSP.}  There is also the possibility that adjoints or bachelors are long lived or even stable \cite{Fox:2002bu}.  This is because these exotics are stable unless additional interactions for them are introduced.  If these come from GUT suppressed operators in the K\"ahler potential, these fields can have lifetimes of seconds and thus will be stable on collider scales.  However, for the adjoints,  an alternative decay is possible if supersoft breaking couples to the visible sector.  As long as this is suppressed compared to the twin sector, EWSB remains natural, while allowing the adjoints to decay promptly into SM particles.  Therefore, there can be a large variety in the exotic sector in their appearance in events.

Some phenomenological aspects differ as the model interpolates between the twin Higgs and SUSY limits, i.e. as $f/m_{\tilde{t}}$ increases.  One interesting signature of the twin Higgs limit is that the Higgs quartic coupling is smaller than expected, since the quartic Higgs coupling stops running above the mass of the $t'$ and not at $m_{\tilde{t}}$.  Also, the radiative correction to the Higgs mass parameter is cut off by the $t'$ mass and not the stop's, giving another potential handle on the twin sector.  If this could be differentiated from an NMSSM or other MSSM extensions, this would be a curious hint of the new physics.  On the other hand, this scenario would be very difficult to fit to a normal SUSY model.  In the SUSY limit, one could interpret it as an MSSM/NMSSM model, but would have only hints of some UV structure that would be interpreted as a particular SUSY breaking scenario.  In the intermediate case between SUSY and twin Higgs, the Higgses would be expected to mix more, leading to larger invisible decays for some of the Higgses, but this would probably be difficult to disentangle from a normal SUSY model.             

In the best possible scenario, precision measurements of the SUSY/Higgs spectra and decays would give a compelling argument for this model.  In this regard, it is fortunate that the naturalness constraints of the model suggest that SUSY phenomenology can be analyzed well at the LHC/ILC.  This is because soft masses for electroweak particles are roughly bounded by 400 GeV and possibly lighter, giving many light sleptons, charginos and neutralinos.  However, unfortunately the mass of the exotics are not guaranteed to be light, so this more distinctive signature is not guaranteed.

\subsection*{A Gauge Mediated Example}
To give a specific example spectrum, we can specialize to the case of a low scale gauge mediation model, with one set of messengers, under the assumption of $SU(3)^3$ unification.  We choose $\Lambda = 60 \,\tev, M_{mess}= 600 \,\tev$ to determine the soft breaking masses and for the bachelors and adjoints we assume a unified $\mu$-term at the GUT scale of the value $\mu_{GUT} = 20 \,\gev$.  With these parameters, we get the spectrum in Table \ref{table:spectrum}.  To avoid complications in the Higgs sector, we do not attempt to break the spectrum down the electroweak gauginos and higgsinos down to physical neutralinos and charginos, but only give some mass entries to give an idea about the rough mass scales.  From the point of view of naturalness, the large contributions to the scalar $T$ mass, suggest that the supersoft scale $m_T$ is quite large, which through unification suggests that the color supersoft scale $m_O$ gives too large a $Z_2$ violation to the twin top squark soft masses.  So a realistic version of this spectrum would require that the coupling to the supersoft sector is not unified.  

The phenomenology of this particular example is much like a normal gauge mediated scenario with the addition of the exotic adjoints and bachelors.  There are decent prospects for discovering the electroweak charged adjoints given their low masses (via judicious choice of $\mu_{GUT}$).  Unfortunately, unification pushes the colored adjoint to be very heavy and perhaps too difficult to discover.  The normal SUSY spectrum is also quite reasonable and would be well explored at the LHC.  Of course, the details of the decays of exotics could have an important impact on how much and how well the details of this model could be mapped out.  In fact, it would be interesting to se if this can be done realistically, in this model or any other specific assumption of the SUSY breaking.         

So far, we have unfortunately not determined a direct signal of the hidden twin sector in this model.  This in fact was one of the motivations for the original Twin Higgs model, in demonstrating that the new physics that made EWSB natural did not have to be charged under the SM.  In this model, the prospects are somewhat intermediate, where measurements within the SUSY phenomenology might give indirect signals of the hidden sector.  The model is even better in that naturalness points to parameter space where the SUSY phenomenology is easily explorable at future colliders.  So even though this is just one possible UV completion of the Twin Higgs, it does demonstrate that new physics signals may still exist in a Twin Higgs realization, maybe even enough to (indirectly) see the effects of a twin sector.    

\begin{table}
\begin{center}
\begin{tabular}{|c|c|c|c|c|c|}
\hline
Scalars & Mass (GeV) & Gaugino & Mass (GeV) & Exotics & Mass (GeV)\\
\hline
$\tilde{e}_1$ & 120 & $M_1$ & 120 & $S,\Psi_S$& 20\\
$\tilde{e}_2$ & 240 & $M_2$& 240 & $T$& 400 \\
$\tilde{\nu}$ & 230 & $M_3$ & 570 & $\Psi_T$& 110\\
$\tilde{q}$ & 930 & & & $O$& 3900\\
& & & & $\Psi_O$ & 3600\\
\hline
\end{tabular}
\end{center}
\caption{Sample spectrum for a single messenger gauge mediation with $\Lambda = 60 \,\tev, \mu_{GUT} = 20 \,\gev$.\label{table:spectrum}}
\end{table}

\section{Discussion}
Supersymmetry, as a general idea, provides an elegant solution to the
hierarchy problem. In practice, it is beset by a number of difficulties. The
radiative electroweak symmetry breaking - an appealing aspect of the MSSM
originally - is now too strong a force, inducing large values of $m_Z$
in the absence of significant fine tuning in the theory.

To this end, we have described a realization of the ``twin Higgs" mechanism
within the context of supersymmetry. Here, the visible sector is related to
a ``twin" sector by a $Z_2$ symmetry.  The Higgs is a pseudo-Goldstone of an
approximate $SU(4)$ symmetry, which arises as an accidental consequence of
the $Z_2$, and the theory cancels the large Higgs mass terms without the
inclusion of any new colored particles. Because the $Z_2$ is only broken by
supersoft supersymmetry breaking operators, contributions to the Higgs mass
are generally loop suppressed, but it maintains the usual $D$-term quartics
of the MSSM.

The only model independent predictions involve some small mixing with the
twin Higgs field, and the existence of an $SU(2)$-triplet. However, there is
a great deal of phenomenology which is expected. GUT-completing the theory
into trinification suggests the presence of a number of exotic fields, which
should be light (100-300 \gev) by naturalness arguments. Such fields may lie
at the end of new cascade decays, and may be stable on collider timescales.
Similarly, we expect light winos, binos and sleptons, as their masses are
indirectly tied to the effectiveness of the twin Higgs mechanism. The
Higgsino mass, $\mu$ is essentially unrelated to the Higgs mass, and so can
be much lighter than in many models.

The presence of light degrees of freedom (new photons and neutrinos) can be
problematic for BBN and the CMBR. However, the $Z_2$ violating supersoft
operators can yield apparently hard $Z_2$ violating terms in the low energy
theory, such as larger Yukawas in the twin sector. Such small changes easily
address the cosmological issues.

This proposal can be easily incorporated into most supersymmetric models,
the most stringent requirements arise from the presence of new fields when
the theory is unified. The most exciting phenomenological consequences arise
from the ``bachelor'' fields - the unmarried GUT partners of the adjoints. A
full discussion of their effects on cascades at the LHC, both with short and
long bachelor lifetimes is warranted.

\acknowledgments
We would like to thank Gia Dvali, Jay Wacker for useful discussions.  The work of S. Chang and N. Weiner was supported by NSF CAREER grant PHY-0449818. The work of L.J. Hall was supported by the US Department of Energy under contracts DE-AC03-76SF00098 and DE-FG03-91ER-40676 and by the National Science Foundation under grant PHY-00-98840.

\bibliographystyle{JHEP}
\bibliography{twin}

\providecommand{\href}[2]{#2}\begingroup\raggedright\begin{thebibliography}{10}

\bibitem{Chacko:2005ra}
Z.~Chacko, Y.~Nomura, and D.~Tucker-Smith, {\it A minimally fine-tuned
  supersymmetric standard model},  {\em Nucl. Phys.} {\bf B725} (2005)
  207--250, [\href{http://xxx.lanl.gov/abs/hep-ph/0504095}{{\tt
  hep-ph/0504095}}].

\bibitem{Choi:2005hd}
K.~Choi, K.~S. Jeong, T.~Kobayashi, and K.-i. Okumura, {\it Little susy
  hierarchy in mixed modulus-anomaly mediation},  {\em Phys. Lett.} {\bf B633}
  (2006) 355--361, [\href{http://xxx.lanl.gov/abs/hep-ph/0508029}{{\tt
  hep-ph/0508029}}].

\bibitem{Kitano:2005wc}
R.~Kitano and Y.~Nomura, {\it A solution to the supersymmetric fine-tuning
  problem within the mssm},  {\em Phys. Lett.} {\bf B631} (2005) 58--67,
  [\href{http://xxx.lanl.gov/abs/hep-ph/0509039}{{\tt hep-ph/0509039}}].

\bibitem{Kobayashi:2004pu}
T.~Kobayashi and H.~Terao, {\it Suppressed supersymmetry breaking terms in the
  higgs sector},  {\em JHEP} {\bf 07} (2004) 026,
  [\href{http://xxx.lanl.gov/abs/hep-ph/0403298}{{\tt hep-ph/0403298}}].

\bibitem{Kobayashi:2005mg}
T.~Kobayashi, H.~Nakano, and H.~Terao, {\it Induced top yukawa coupling and
  suppressed higgs mass parameters},  {\em Phys. Rev.} {\bf D71} (2005) 115009,
  [\href{http://xxx.lanl.gov/abs/hep-ph/0502006}{{\tt hep-ph/0502006}}].

\bibitem{Birkedal:2004xi}
A.~Birkedal, Z.~Chacko, and M.~K. Gaillard, {\it Little supersymmetry and the
  supersymmetric little hierarchy problem},  {\em JHEP} {\bf 10} (2004) 036,
  [\href{http://xxx.lanl.gov/abs/hep-ph/0404197}{{\tt hep-ph/0404197}}].

\bibitem{Berezhiani:2005pb}
Z.~Berezhiani, P.~H. Chankowski, A.~Falkowski, and S.~Pokorski, {\it Double
  protection of the higgs potential in a supersymmetric little higgs model},
  {\em Phys. Rev. Lett.} {\bf 96} (2006) 031801,
  [\href{http://xxx.lanl.gov/abs/hep-ph/0509311}{{\tt hep-ph/0509311}}].

\bibitem{Roy:2005hg}
T.~Roy and M.~Schmaltz, {\it Naturally heavy superpartners and a little higgs},
   {\em JHEP} {\bf 01} (2006) 149,
  [\href{http://xxx.lanl.gov/abs/hep-ph/0509357}{{\tt hep-ph/0509357}}].

\bibitem{Csaki:2005fc}
C.~Csaki, G.~Marandella, Y.~Shirman, and A.~Strumia, {\it The super-little
  higgs},  {\em Phys. Rev.} {\bf D73} (2006) 035006,
  [\href{http://xxx.lanl.gov/abs/hep-ph/0510294}{{\tt hep-ph/0510294}}].

\bibitem{Harnik:2003rs}
R.~Harnik, G.~D. Kribs, D.~T. Larson, and H.~Murayama, {\it The minimal
  supersymmetric fat higgs model},  {\em Phys. Rev.} {\bf D70} (2004) 015002,
  [\href{http://xxx.lanl.gov/abs/hep-ph/0311349}{{\tt hep-ph/0311349}}].

\bibitem{Chang:2004db}
S.~Chang, C.~Kilic, and R.~Mahbubani, {\it The new fat higgs: Slimmer and more
  attractive},  {\em Phys. Rev.} {\bf D71} (2005) 015003,
  [\href{http://xxx.lanl.gov/abs/hep-ph/0405267}{{\tt hep-ph/0405267}}].

\bibitem{Delgado:2005fq}
A.~Delgado, and Tim M.~P.~Tait, {\it A fat Higgs with a fat top},  {\em JHEP} {\bf 07} (2005),
  [\href{http://xxx.lanl.gov/abs/hep-ph/0504224}{{\tt hep-ph/0504224}}].



\bibitem{Batra:2003nj}
P.~Batra, A.~Delgado, D.~E. Kaplan, and T.~M.~P. Tait, {\it The higgs mass
  bound in gauge extensions of the minimal supersymmetric standard model},
  {\em JHEP} {\bf 02} (2004) 043,
  [\href{http://xxx.lanl.gov/abs/hep-ph/0309149}{{\tt hep-ph/0309149}}].

\bibitem{Maloney:2004rc}
A.~Maloney, A.~Pierce, and J.~G. Wacker, {\it D-terms, unification, and the
  higgs mass},  \href{http://xxx.lanl.gov/abs/hep-ph/0409127}{{\tt
  hep-ph/0409127}}.

\bibitem{Espinosa:1992hp}
J.~R. Espinosa and M.~Quiros, {\it Upper bounds on the lightest higgs boson
  mass in general supersymmetric standard models},  {\em Phys. Lett.} {\bf
  B302} (1993) 51--58, [\href{http://xxx.lanl.gov/abs/hep-ph/9212305}{{\tt
  hep-ph/9212305}}].

\bibitem{Dermisek:2005ar}
R.~Dermisek and J.~F. Gunion, {\it Escaping the large fine tuning and little
  hierarchy problems in the next to minimal supersymmetric model and h --> a a
  decays},  {\em Phys. Rev. Lett.} {\bf 95} (2005) 041801,
  [\href{http://xxx.lanl.gov/abs/hep-ph/0502105}{{\tt hep-ph/0502105}}].

\bibitem{Chang:2005ht}
S.~Chang, P.~J. Fox, and N.~Weiner, {\it Naturalness and higgs decays in the
  mssm with a singlet},  \href{http://xxx.lanl.gov/abs/hep-ph/0511250}{{\tt
  hep-ph/0511250}}.

\bibitem{Schuster:2005py}
P.~C. Schuster and N.~Toro, {\it Persistent fine-tuning in supersymmetry and
  the nmssm},  \href{http://xxx.lanl.gov/abs/hep-ph/0512189}{{\tt
  hep-ph/0512189}}.

\bibitem{Dermisek:2005gg}
R.~Dermisek and J.~F. Gunion, {\it Consistency of lep event excesses with an h
  --> a a decay scenario and low-fine-tuning nmssm models},
  \href{http://xxx.lanl.gov/abs/hep-ph/0510322 v4}{{\tt hep-ph/0510322 v4}}.

\bibitem{Chacko:2005pe}
Z.~Chacko, H.-S. Goh, and R.~Harnik, {\it The twin higgs: Natural electroweak
  breaking from mirror symmetry},
  \href{http://xxx.lanl.gov/abs/hep-ph/0506256}{{\tt hep-ph/0506256}}.

\bibitem{Barbieri:2005ri}
R.~Barbieri, T.~Gregoire, and L.~J. Hall, {\it Mirror world at the large hadron
  collider},  \href{http://xxx.lanl.gov/abs/hep-ph/0509242}{{\tt
  hep-ph/0509242}}.

\bibitem{Chacko:2005vw}
Z.~Chacko, Y.~Nomura, M.~Papucci, and G.~Perez, {\it Natural little hierarchy
  from a partially goldstone twin higgs},  {\em JHEP} {\bf 01} (2006) 126,
  [\href{http://xxx.lanl.gov/abs/hep-ph/0510273}{{\tt hep-ph/0510273}}].

\bibitem{Chacko:2005un}
Z.~Chacko, H.-S. Goh, and R.~Harnik, {\it A twin higgs model from left-right
  symmetry},  {\em JHEP} {\bf 01} (2006) 108,
  [\href{http://xxx.lanl.gov/abs/hep-ph/0512088}{{\tt hep-ph/0512088}}].

\bibitem{Fox:2002bu}
P.~J. Fox, A.~E. Nelson, and N.~Weiner, {\it Dirac gaugino masses and supersoft
  supersymmetry breaking},  {\em JHEP} {\bf 08} (2002) 035,
  [\href{http://xxx.lanl.gov/abs/hep-ph/0206096}{{\tt hep-ph/0206096}}].

\bibitem{martin}
A.~Falkowski, S.~Pokorski, and M.~Schmaltz, {\it Twin susy},
 \href{http://xxx.lanl.gov/abs/hep-ph/0604066}{{\tt hep-ph/0604066}}.


\bibitem{Martin:1997ns}
S.~P. Martin, {\it A supersymmetry primer},
  \href{http://xxx.lanl.gov/abs/hep-ph/9709356}{{\tt hep-ph/9709356}}.

\bibitem{Arkani-Hamed:1998kj}
N.~Arkani-Hamed, G.~F. Giudice, M.~A. Luty, and R.~Rattazzi, {\it
  Supersymmetry-breaking loops from analytic continuation into superspace},
  {\em Phys. Rev.} {\bf D58} (1998) 115005,
  [\href{http://xxx.lanl.gov/abs/hep-ph/9803290}{{\tt hep-ph/9803290}}].

\bibitem{Kaplan:1999ac}
D.~E. Kaplan, G.~D. Kribs, and M.~Schmaltz, {\it Supersymmetry breaking through
  transparent extra dimensions},  {\em Phys. Rev.} {\bf D62} (2000) 035010,
  [\href{http://xxx.lanl.gov/abs/hep-ph/9911293}{{\tt hep-ph/9911293}}].

\bibitem{Chacko:1999mi}
Z.~Chacko, M.~A. Luty, A.~E. Nelson, and E.~Ponton, {\it Gaugino mediated
  supersymmetry breaking},  {\em JHEP} {\bf 01} (2000) 003,
  [\href{http://xxx.lanl.gov/abs/hep-ph/9911323}{{\tt hep-ph/9911323}}].

\bibitem{Froggatt:1978nt}
C.~D. Froggatt and H.~B. Nielsen, {\it Hierarchy of quark masses, cabibbo
  angles and cp violation},  {\em Nucl. Phys.} {\bf B147} (1979) 277.

\bibitem{Dolgov:2003sg}
A.~D. Dolgov and F.~L. Villante, {\it Bbn bounds on active-sterile neutrino
  mixing},  {\em Nucl. Phys.} {\bf B679} (2004) 261--298,
  [\href{http://xxx.lanl.gov/abs/hep-ph/0308083}{{\tt hep-ph/0308083}}].

\bibitem{Schabinger:2005ei}
R.~Schabinger and J.~D. Wells, {\it A minimal spontaneously broken hidden
  sector and its impact on higgs boson physics at the large hadron collider},
  {\em Phys. Rev.} {\bf D72} (2005) 093007,
  [\href{http://xxx.lanl.gov/abs/hep-ph/0509209}{{\tt hep-ph/0509209}}].

\end{thebibliography}\endgroup

\end{document}